\documentclass[eqsecnum,nofootinbib,11pt]{revtex4}
\usepackage{amsmath}
   \usepackage{bm}
\usepackage{amsthm,amscd}
\usepackage{mathrsfs}
\usepackage{verbatim}
\usepackage{amsfonts,amsmath,latexsym,amssymb,bm}
\usepackage[american]{babel}
\usepackage{dsfont}
\usepackage[dvips]{graphicx}

\newcommand\bea{\begin{eqnarray}}
\newcommand\eea{\end{eqnarray}}

\begin{document}
\title{Heat Kernel Coefficients for Laplace Operators on the Spherical Suspension}
\author{
Guglielmo Fucci\footnote{Electronic address: Guglielmo\textunderscore Fucci@Baylor.edu} and Klaus Kirsten\footnote{Electronic address: Klaus\textunderscore Kirsten@Baylor.edu}
\thanks{Electronic address: gfucci@nmt.edu}}
%\altaffiliation[Also at ]{Department of Mathematics, Baylor University One Bear Place 9653, Waco, TX USA}
\affiliation{Department of Mathematics, Baylor University, Waco, TX 76798 USA
}
\date{\today}
\vspace{2cm}
\begin{abstract}

In this paper we compute the coefficients of the heat kernel asymptotic expansion for Laplace operators acting on scalar functions
defined on the so called spherical suspension (or Riemann cap) subjected to Dirichlet boundary conditions.
By utilizing a contour integral representation of the spectral zeta function for the Laplacian on the spherical suspension we find its analytic continuation in the complex plane and its associated meromorphic structure.
Thanks to the well known relation between the zeta function and the heat kernel obtainable via Mellin transform
we compute the coefficients of the asymptotic expansion in arbitrary dimensions. The particular case of a $d$-dimensional sphere as the
base manifold is studied as well and the first few heat kernel coefficients are given explicitly.

\end{abstract}
\maketitle

\section{Introduction}

The study of the heat kernel for Laplace type operators has been a topic of major interest in the literature. The reason lies in the
fact that it represents a very versatile tool for the analysis of various problems. Its interest in mathematics
resides, especially, in geometric analysis thanks to the well known relation between the heat equation and the
Atiyah-Singer index theorem \cite{atiyah63,atiyah73,gilkey95}. In physics, the heat kernel of an elliptic
operator and its associated spectral zeta function are particularly useful in the ambit
of quantum field theory in order to compute the one-loop effective action and the Casimir energy
\cite{birrell,cognola92,fulling,hawking77,kirsten01,vassilevich03}.

The computation of the heat kernel, however, is not a trivial task.
In fact, its exact form can only be found for highly symmetric backgrounds when the spectrum of the operator is
explicitly known. For this reason, various approximations have been developed and used throughout the years.
One of the most important approximation schemes is the short time asymptotic expansion
of the heat kernel also known as Schwinger-DeWitt expansion \cite{dewitt65,dewitt67,dewitt67a,dewitt75}.
Very well established methods have been developed in order to compute the coefficients of the heat kernel
asymptotic expansion for elliptic operators on compact manifolds with and without boundary (see e.g. \cite{avramidi91,dewitt75,vassilevich03,gilkey95,gilkey04,kirsten01}).
A very efficient way for evaluating the coefficients on a manifold with boundary relies on the use of the spectral zeta function for special cases. This is done by exploiting
the well known relation between the trace of the heat kernel and the associated spectral zeta function \cite{bordag96,bordag96a,bordag96b,kirsten01}. We provide a new example here that can be used for this purpose.

It is well known that for second order self-adjoint elliptic partial differential operators on compact smooth manifolds
the spectrum $\lambda_n$ is discrete, bounded from below and it forms a monotonically increasing sequence of numbers tending to infinity.
The spectral zeta function for such operators is defined as the following sum
\begin{equation}\label{00}
\zeta(s)=\sum_{n=1}^\infty\lambda_{n}^{-s}\;,
\end{equation}
which is convergent for $\Re(s)> D/2$, with $D$ being the dimension of the manifold under consideration.
One can analytically continue $\zeta(s)$, in a unique way, to a meromorphic function in the whole
complex plane which coincides with (\ref{00}) in its domain of convergence. In this paper we will utilize spectral zeta function
techniques in order to evaluate the coefficients of the heat kernel asymptotic expansion for a Laplace operator, with Dirichlet boundary conditions, on a singular
Riemannian manifold known as the spherical suspension. A related work on the spherical suspension is \cite{flachi10},
where the functional determinant for a Laplace operator and the conformal anomaly have been analyzed.

The current paper represents
somewhat a generalization of the conical geometries studied in \cite{bordag96}. The difference is not in the mathematical
technique used, but in the radial eigenfunctions of the Laplace operator that one finds. In the case of the generalized cone, a prototypical singular Riemannian manifold, and in many other
ones considered in the literature, the radial eigenfunctions are Bessel functions. Well developed methods have been described in order
to use their uniform asymptotic expansion for the analytic continuation of the spectral zeta function.
In the case of the spherical suspension considered here the radial eigenfunctions are associated Legendre functions. To our knowledge,
the analytic continuation of the associated spectral zeta function by making use of the uniform asymptotic expansion
of the Legendre function is analyzed here for the first time. As we will see in the next sections, the technical details
are quite different from the Bessel case which makes this work particularly interesting. It is worth mentioning that
zeta function regularization techniques involving associated Legendre functions have been considered in the study of the one-loop effective action
for gravitons in a DeSitter spacetime \cite{barvinsky}. In that paper, however, a different procedure was used from the one presented in our work.

The layout of the paper is as follows. In the next section we will describe the geometry of the
spherical suspension and present the eigenvalue problem we are interested in. In the subsequent sections we will construct
the zeta function in terms of a contour integral and explicitly find its analytic continuation. This will allow us to
compute, in section \ref{sec5}, the heat kernel coefficients on the spherical suspension in terms of those on its base.
As a particular case in section \ref{sec6}, we study the situation in which the base manifold is a $d$-dimensional sphere which will permit us to find more
explicit results.

\section{The geometry and Heat Kernel}

The manifold that we will consider in this paper is termed spherical suspension or Riemann cap. This is defined as
the $D=(d+1)$-dimensional manifold $\mathscr{M}=\mathcal{I}\times\mathscr{N}$ where $\mathcal{I}\subseteq[0,\theta_{0}]$, with $\theta_0\in(0,\pi)$, and where $\mathscr{N}$
represents a smooth, compact Riemannian $d$-dimensional manifold, possibly with boundary, which will be referred to as the base manifold. The local geometry
of $\mathscr{M}$ can be written in hyperspherical coordinates as \cite{bordag96,flachi10}
\begin{equation}\label{0}
ds^2=d\theta^2 +\sin^2\theta d\Sigma^2\;,
\end{equation}
where $d\Sigma^2$ represents the line element on the base manifold $\mathscr{N}$. It is easily seen, from (\ref{0}), that
the Riemann cap $\mathscr{M}$ in general possesses a singularity and, therefore, it belongs to the class of singular Riemannian
manifolds. Since $\sin\theta_0\sim\theta_0$ in the limit of small angles $\theta_0$ the metric (\ref{0}) reduces to the one representing a generalized cone \cite{bordag96}.
This special limiting case will provide a robust check of the results that we will obtain in this paper.

In hyperspherical coordinates the Laplacian on $\mathscr{M}$, namely $\Delta_{\mathscr{M}}$, can be written as follows
\begin{equation}\label{1}
\Delta_\mathscr{M} =
{\partial^2 \over \partial \theta^2}
+d \cot \theta {\partial\over \partial \theta}
+{1\over \sin^2\theta}\Delta_\mathscr{N}\;,
\end{equation}
where $\Delta_\mathscr{N}$ is identified with the Laplacian on the base manifold $\mathscr{N}$. The eigenvalue equation
\begin{equation}\label{2}
-\Delta_{\mathscr{M}}\varphi=\alpha^{2}\varphi\;,
\end{equation}
with $\varphi$ belonging to the space $\mathcal{L}^{2}(\mathscr{M})$,
is separable in this system of coordinates and a general solution to (\ref{2}) can be written
as a product
\begin{equation}\label{3}
\varphi =
(\sin\theta)^{\frac{(1-d)}{2}} \psi(\theta){\cal H}\;.
\end{equation}
Here, the functions ${\cal H}$ are defined to be the harmonics on $\mathscr{N}$ having degeneracy $d( \lambda )$ and satisfying the following eigenvalue equation
\begin{equation}\label{4}
-\Delta_\mathscr{N} {\cal H}=
\lambda^2
{\cal H}\;.
\end{equation}
As will become clear later, the eigenvalue problem resulting from
\begin{eqnarray}\label{eignew}
\left(-\Delta_{\mathscr{M}} + \frac{d^2} 4  \right) \varphi = \omega^2 \varphi ,
\end{eqnarray}
has certain advantages compared to (\ref{2}). In this case the function
$\psi$ in (\ref{3}) obeys the differential equation of Legendre type
\begin{equation}\label{6}
{d^2\over d\theta^2} +\cot\theta {d\over d\theta}
+ \left[ \left( -{1\over 4}-\omega^2 \right)
-\left(
\lambda^2 + {(d-1)^2\over 4}
\right)
\frac{1}{\sin^{2} \theta}
\right] = 0.
\end{equation}
This is the eigenvalue problem we will concentrate on and we will make remarks at appropriated places on
how the original eigenvalue problem (\ref{2}) can be recovered and what the additional complications are.

Let us next turn our attention to the associated heat kernel and its asymptotic expansion. For explanatory purposes, let $M$ be a smooth manifold with boundary $\partial M$. It is well known that
the $\mathcal{L}^{2}$-trace of the heat kernel has the following small $t$ asymptotic expansion \cite{elizalde,esposito97,gilkey95,kirsten01}
\bea\label{6a}
K(t)\sim\sum_{k=0}^{\infty}E_{\frac{k}{2}}t^{\frac{k-D}{2}}\;,
\eea
where the coefficients
\bea
E_{\frac{k}{2}}=\int_{M}dV\,e_{\frac{k}{2}}+\int_{\partial M}dS\,f_{\frac{k}{2}}\;,
\eea
consist of a volume and a boundary part,
with $e_{\frac{k}{2}}$ and $f_{\frac{k}{2}}$ being polynomials in the local curvature of the manifold. There exists a deep relation between the trace of the heat kernel and the associated spectral zeta function, through the Mellin transform, which is provided by the formulas \cite{elizalde,kirsten01,seel68-10-288}
\bea\label{6b}
E_{\frac{n}{2}-s}=\Gamma(s)\textrm{Res}\,\zeta(s)\;,
\eea
with $s=n/2,(n-1)/2,\cdots,1/2$ and $s=-(2l+1)/2$ for $l\in\mathbb{N}_{0}$,
\bea\label{6c}
E_{\frac{n}{2}+p}=\frac{(-1)^{p}}{p!}\zeta(-p)\;,
\eea
where $p\in\mathbb{N}_{0}$. These relations produce any coefficient of the heat kernel asymptotic expansion in terms of either the residue
or the value of the associated zeta function at a specific point and they will be used in this paper for the computation of the heat kernel
coefficients on the spherical suspension. It is clear that the heat kernels associated with the eigenvalue problems (\ref{2}) and (\ref{eignew}) are related by a factor of $e^{-d^2t/4}$. So,
as far as the heat kernel is concerned, considering (\ref{eignew}) is as informative as considering (\ref{2}). The corresponding changes in the heat kernel coefficients can be trivially obtained.
However, for a full analysis of the associated zeta functions the presence of a mass term $d^2/4$ would cause major complications and we will further comment on these later.

%As a last introductory remark we would like to point out an important property enjoyed by the trace of the heat kernel.
%If $Q$ represents a constant endomorphism and $\mathscr{L}$ a differential operator, then the $\mathcal{L}^{2}$-trace of the heat semigroup possesses the factorization property
%\begin{equation}\label{7}
%\textrm{Tr}_{\mathcal{L}^{2}}e^{-t(\mathscr{L}+Q)}=e^{-tQ}\;\textrm{Tr}_{\mathcal{L}^{2}}e^{-t\mathscr{L}}\;.
%\end{equation}
%This property, specialized to our case, allows us to write that
%\begin{equation}\label{8}
%\textrm{Tr}_{\mathcal{L}^{2}}e^{t(\Delta_{\mathscr{M}}-m^{2})}=e^{-t\left(m^{2}-\frac{d^{2}}{4}\right)}\;\textrm{Tr}_{\mathcal{L}^{2}}e^{-t\mathscr{L}_{m=0}}\;.
%\end{equation}
%From the last formula it is not difficult to realize that in order to compute the heat kernel asymptotic coefficients on the
%$D$-dimensional Riemann cap, it will be sufficient to study the asymptotics of the heat trace for the operator $\mathscr{L}_{m=0}$. For future convenience we
%will consider the spectral zeta function of the massive operator $\mathscr{L}_{m}$ and we will set $m=0$ only in the final result.
%The reason for proceeding in this way is twofold; the presence of the mass parameter will allow for an easier analytic continuation of the relevant integrals,
%and moreover, we will obtain an expression for the spectral zeta function which is valid also when a small, non-vanishing mass is present.

\section{The Spectral $\zeta$-function}

Let us now begin to study the spectral zeta function of the eigenvalue problem (\ref{eignew}).
By introducing
\bea\label{9}
\mu&=&\sqrt{\lambda^2+\frac{(d-1)^{2}}{4}}~,
\eea
a solution $\phi$ to the differential equation (\ref{eignew}) which is regular for $\theta\to 0$ is \cite{flachi10}
\bea\label{10}
\psi (\theta )=\mbox{P}_{-{1\over 2}+ \omega}^{- \mu}\left(\cos\theta\right)\;,
\eea
where $\mbox{P}_{-{1/ 2}+ \omega}^{- \mu}$ corresponds to the {\it Ferrers} representation of the Legendre function \cite{olver}.

The spectral zeta function for the problem under consideration is defined as follows
\bea\label{11}
\zeta(s)=\sum_{\omega} \omega^{-2s}\;,
\eea
where we will assume that no negative eigenvalues occur so that we can use the standard branch cut of the logarithm.
Since $\mathscr{N}$ for now is unspecified, we will express the spectral zeta function on the whole manifold
in terms of the zeta function $\zeta_{\mathscr{N}}(s)$ associated with the operator $-\Delta_{\mathscr{N}}+(d-1)^{2}/4$. Its definition is \cite{cheeger83}
\bea\label{12}
\zeta_{\mathscr{N}}(s)=\sum_{\mu} d(\mu)\mu^{-2s}\;.
\eea
Although the base manifold has not been explicitly chosen, one is still able to impose boundary conditions.
For definiteness we will impose Dirichlet
boundary conditions on the solution (\ref{10}) at $\theta=\theta_{0}$. By doing so, we are led to the relation
\bea\label{13}
\mbox{P}_{-1/2+\omega}^{-\mu} \left(\cos \theta_0\right)=0\;.
\eea
The last equation determines implicitly the eigenvalues $\omega^2$.

It is important to make a remark at this point.
It is well known \cite{bruning84,bruning87,kirsten01} that for singular problems, such as the one considered in this paper,
the heat kernel asymptotic expansion possesses a non-standard logarithmic term. This means that instead of a small $t$ asymptotic expansion of the form  (\ref{6a})
one has
\bea
\mathcal{K}(t)\sim\sum_{k=0}^{\infty}\mathcal{A}^{\mathscr{M}}_{\frac{k}{2}}t^{\frac{k-D}{2}}+\mathscr{G}\ln\,t\;.
\eea
This unusual behavior of the heat kernel leads to the appearance of a pole in the spectral zeta function of $\mathscr{M}$ at $s=0$
and, in addition, one can prove that $\mathscr{G}=-\textrm{Res}\,\zeta(0)$ \cite{bordag96}.
It has been shown in \cite{flachi10} that for the spherical suspension the following relation holds
\bea
\mathscr{G}=\frac{1}{2}\textrm{Res}\,\zeta_{{\mathscr{N}}}\left(-\frac{1}{2}\right)=-\textrm{Res}\,\zeta(0)\;,
\eea
which is the same that one obtains in the framework of the generalized cone \cite{bordag96}. It is clear that
the coefficient of the logarithmic term in the heat kernel asymptotic expansion can be explicitly evaluated
once the residue of $\zeta_{\mathscr{N}}(s)$ at $s=-1/2$ is known.

In order to compute the spectral zeta function, which will be used to evaluate the heat kernel coefficients, we will temporarily consider the massive case. This means we will add a constant
term $m^2$ to the eigenvalues $\omega^2$, which at a later stage will be sent to zero. The reason for proceeding in this way is twofold. First, the presence of the mass parameter will allow for an easier analytic continuation of relevant integrals. Second, we will obtain an expression for the spectral zeta function which is valid also for a massive scalar field, results useful when Casimir energies for massive fields in the given
setting are considered.

To proceed, we express
$\zeta(s)$ in terms of a contour integral in the complex plane, valid for $\Re(s)>(d+1)/2$ as \cite{bordag96,bordag96a,bordag96b,esposito97,kirsten01},
\bea\label{14}
\zeta(s)=\sum_{\mu}d(\mu)\frac{1}{2\pi i}\int_{\Gamma}dk \left(k^{2}+m^{2}\right)^{-s}\frac{\partial}{\partial k}\ln \mbox{P}_{-1/2+k}^{-\mu} \left(\cos \theta_0\right)\;,
\eea
where $\Gamma$ represents a contour that encircles all the roots of $\mbox{P}_{-1/2+k}^{-\mu} \left(\cos \theta_0\right)$ on the positive real axis in the counterclockwise
direction. By deforming the contour of integration $\Gamma$ along the imaginary axis and by using the property $\mbox{P}^{\pm \mu}_{\nu}=\mbox{P}^{\pm \mu}_{-1-\nu}$ \cite{erdelyi53,olver}, the expression (\ref{14}) can be rewritten in the following form
\bea\label{15}
\zeta(s)=\sum_{\mu} d(\mu) \frac{\sin\pi s}{\pi}\int_{m}^{\infty}dz \left(z^{2}-m^{2}\right)^{-s}\frac{\partial}{\partial z}\ln P^{-\mu}_{-\frac{1}{2}+ i z}(\cos\theta_{0})\;.
\eea
Here, the integral
is well defined in the strip $1/2<\Re(s)<1$ and $\zeta (s)$ above has to be interpreted as the function resulting from analytic continuation out of that strip.
By conveniently changing the integration variable $z\to \mu w/\sin\theta_0$, one can rewrite
(\ref{15}) as
\bea\label{16}
\zeta(s)=\sum_{\mu} d(\mu) \frac{\sin\pi s}{\pi}\int_{\frac{m}{\mu}\sin\theta_0}^{\infty}dw \left(\frac{\mu^{2}w^{2}}{\sin^{2}\theta_0}-m^{2}\right)^{-s}\frac{\partial}{\partial w}\ln P^{-\mu}_{-\frac{1}{2}+i \frac{\mu w}{\sin\theta_0}}(\cos\theta_{0})\;.
\eea

By following the technique developed in \cite{bordag96,bordag96a,bordag96b}, we now utilize the uniform asymptotic expansion
of the Legendre functions for $\mu\to\infty$. By exploiting the WKB method, one obtains the following uniform asymptotic expansion valid for $\mu\to\infty$ and fixed $w$
\cite{flachi10,khusnutdinov03,thorne57}
\bea\label{17}
P^{-\mu}_{-\frac{1}{2}+i\frac{ \mu w}{\sin\theta_{0}}}(\cos\theta_{0})\sim\sqrt{\frac{t}{2\pi\mu}}\;e^{\mu\eta}\left(\frac{\sin\theta_{0}}{w\mu}\right)^{\mu}\sum_{l=0}^{\infty}\frac{\Psi_{l}(\nu)}{\mu^{l}}\;,
\eea
where we have introduced the notation
\bea\label{18}
t&=&\frac{1}{\sqrt{1+w^{2}}}\;,\quad \nu=\frac{\cos\theta_{0}}{\sqrt{1+w^{2}}}\;,
\eea
and
\bea\label{19}
\eta&=&1+\ln\left[\frac{w}{\cos\theta_{0}+\sqrt{1+w^{2}}}\right]-\frac{w}{\sin\theta_{0}}\left[\arctan\left(\frac{\sin\theta_{0}}{w}\right)-\arctan\left(\frac{\tan\theta_{0}}{t w}\right)\right]\;.
\eea
The functions $\Psi_{l}(\nu)$ in expression (\ref{17}) are defined by the formula
\bea\label{20}
\sum_{n=0}^{\infty}\mu^{-n}\Psi_{n}(\nu)=\exp\left\{-\sum_{l=1}^{\infty}\frac{B_{2l}}{2l(2l-1)\mu^{2l-1}}\right\}\sum_{j=0}^{\infty}\mu^{-j}\Phi_{j}(\nu)\;,
\eea
where $B_{2l}$ are the Bernoulli numbers and the functions $\Phi_{n}(\nu)$ satisfy the recurrence relation \cite{flachi10,khusnutdinov03}
\bea\label{21}
\Phi_{n+1}(\nu)&=&\frac{(1-\nu^{2})(1+\gamma^{2}\nu^{2})}{2(1+\gamma^{2})}\partial_{\nu}\left[\Phi_{n}(\nu)\right]
-\frac{\gamma^{2}}{8(1+\gamma^{2})}\int_{1}^{\nu}d\nu^{\prime}\left[5\nu'+\frac{1}{\gamma^{2}}-1\right]\Phi_{n}(\nu')\;,
\eea
with $\Phi_{0}=1$ and $\gamma=w/\sin\theta_0$. For the purpose of this paper we will actually need the uniform asymptotic expansion for the
logarithm of the Legendre function (\ref{17}). To this end, it is straightforward to obtain
\bea\label{22}
\ln P^{-\mu}_{-\frac{1}{2}+i\frac{ \mu w}{\sin\theta_{0}}}(\cos\theta_{0})\sim\ln\left\{\sqrt{\frac{t}{2\pi\mu}}\;e^{\mu\eta}\left(\frac{\sin\theta_{0}}{w\mu}\right)^{\mu}\right\}+\sum_{n=1}^{\infty}\frac{\Omega_{n}(\nu)}{\mu^{n}}\;,
\eea
with the newly introduced functions $\Omega_{n}(\nu)$ being defined through the cumulant expansion
\bea\label{23}
-\sum_{l=1}^{\infty}\frac{B_{2l}}{2l(2l-1)\mu^{2l-1}}+\ln\left\{1+\sum_{j=1}^{\infty}\mu^{-j}\Phi_{j}(\nu)\right\}\sim\sum_{n=1}^{\infty}\frac{\Omega_{n}(\nu)}{\mu^{n}}\;.
\eea
The first few $\Omega_{n}(\nu)$ are listed for completeness in appendix \ref{app1}.

By adding and subtracting $N$ leading terms of the asymptotic expansion (\ref{22}), one can write the spectral zeta function $\zeta(s)$
in the following way (cf. \cite{bordag96,bordag96a,bordag96b,kirsten01})
\bea\label{24}
\zeta(s)=Z(s)+\sum_{i=-1}^{N}A_{i}(s)\;,
\eea
where the terms $Z(s)$ and $A_{i}(s)$ are defined as
\bea\label{25}
Z(s)&=&\sum_{\mu}d(\mu)\frac{\sin\pi s}{\pi}\int_{\frac{m}{\mu}\sin\theta_{0}}^{\infty}dw\left[\frac{\mu^{2}w^{2}}{\sin^{2}\theta_{0}}-m^{2}\right]^{-s}\frac{\partial}{\partial w}
\Bigg\{\ln\left[ P^{-\mu}_{-\frac{1}{2}+i\frac{ \mu w}{\sin\theta_{0}}}(\cos\theta_{0})\right]\nonumber\\
&-&\ln\left[\sqrt{\frac{t}{2\pi\mu}}e^{\mu\eta}\left(\frac{\sin\theta_{0}}{\mu w}\right)^{\mu}\right]-\sum_{n=1}^{N}\frac{\Omega_{n}(\nu)}{\mu^{n}}
\Bigg\}\;,
\eea
\bea\label{26}
A_{-1}(s)=\sum_{\mu}d(\mu)\frac{\sin\pi s}{\pi}\int_{\frac{m}{\mu}\sin\theta_{0}}^{\infty}dw\left[\frac{\mu^{2}w^{2}}{\sin^{2}\theta_{0}}-m^{2}\right]^{-s}
\frac{\partial}{\partial w}\left\{\ln\left[\left(\frac{\sin\theta_{0}}{\mu w}\right)^{\mu}e^{\mu\eta}\right]\right\}\;,
\eea
\bea\label{27}
A_{0}(s)=\sum_{\mu}d(\mu)\frac{\sin\pi s}{\pi}\int_{\frac{m}{\mu}\sin\theta_{0}}^{\infty}dw\left[\frac{\mu^{2}w^{2}}{\sin^{2}\theta_{0}}-m^{2}\right]^{-s}\frac{\partial}{\partial w}\left\{\ln\left[\sqrt{\frac{t}{2\pi\mu}}\right]\right\}\;,
\eea
and finally
\bea\label{28}
A_{i}(s)=\sum_{\mu}d(\mu)\frac{\sin\pi s}{\pi}\int_{\frac{m}{\mu}\sin\theta_{0}}^{\infty}dw\left[\frac{\mu^{2}w^{2}}{\sin^{2}\theta_{0}}-m^{2}\right]^{-s}\frac{\partial}{\partial w}\left\{\frac{\Omega_{i}(\nu)}{\mu^{i}}\right\}\;.
\eea
It is very important to notice that $Z(s)$ is an analytic function in the region $\Re(s)>(d-1-N)/2$. This restriction is found by considering the behavior
on the integrand in $\mu$, which is $\mu^{-2s-N-1}$ and by recalling that the spectral zeta function on $\mathscr{N}$ converges for $\Re(s)>d/2$ according to Weyl's estimate. This means that in any dimension $D$ one can choose a suitable $N$ such that $Z(s)$ does not contribute to the residue of the zeta function $\zeta(s)$ in the region $\Re(s)>(d-1-N)/2$, and hence to the value of the heat kernel coefficients (cf. \ref{6b}) \cite{bordag96a,kirsten01}.

\section{Evaluation of $A_{-1}(s)$, $A_{0}(s)$ and $A_{i}(s)$}

In this section we are mainly interested in the non trivial task of explicitly evaluating the functions in (\ref{26})-(\ref{28}). According to the remarks made earlier, these are the only terms that
will contribute to the coefficients of the heat kernel expansion. Let us start with the leading term $A_{-1}(s)$. By recalling the expression for $\eta$ in (\ref{19}) one obtains
\bea\label{29}
A_{-1}(s)&=&\sum_{\mu}d(\mu)\mu\frac{\sin\pi s}{\pi\sin\theta_0}\int_{\frac{m}{\mu}\sin\theta_{0}}^{\infty}dw\left[\frac{\mu^{2}w^{2}}{\sin^{2}\theta_{0}}-m^{2}\right]^{-s}\Bigg\{
\arctan\left(\frac{\sqrt{1+w^{2}}}{w}\tan\theta_{0}\right)\nonumber\\
&-&\arctan\left(\frac{\sin\theta_{0}}{w}\right)
\Bigg\}\;.
\eea
It does not seem possible to find a simple analytic expression for the above integral, however some progress can be made by
considering the following expansion
\bea\label{30}
& &\frac{1}{\sin\theta_0}\left[\arctan\left(\frac{\sqrt{1+w^{2}}}{w}\tan\theta_{0}\right)-
\arctan\left(\frac{\sin\theta_{0}}{w}\right)\right]=\sum_{k=0}^{\infty}\frac{(2k)!}{2^{2k}(k!)^{2}(2k+1)\Gamma\left(k+\frac{1}{2}\right)}(\sin\theta_0)^{2k}\nonumber\\
& &\times\sum_{j=0}^{\infty}\frac{(-1)^{j}}{j!}\Gamma\left(j+k+\frac{1}{2}\right)(\sin\theta_0)^{2j}w^{-2j}\left[\left(1+\frac{1}{w^{2}}\right)^{k+\frac{1}{2}}-\frac{1}{w^{2k+1}}\right]\;.
\eea
The expansion above is obtained by first using the relation \cite{gradshtein07}
\bea
\arctan\,x=\arcsin\frac{x}{\sqrt{1+x^{2}}}\;,
\eea
and then by expanding the arcsine for small arguments.
By using the result (\ref{30}) in the integral (\ref{29}), and by making the change of variable $w\to m\sin\theta_0/(\mu t)$ one obtains
\bea\label{31}
A_{-1}(s)&=&\sum_{\mu}d(\mu)\frac{\sin\pi s}{\pi}m^{-2s+1}\sum_{k=0}^{\infty}\frac{(2k)!}{2^{2k}(k!)^{2}(2k+1)\Gamma\left(k+\frac{1}{2}\right)}\sum_{j=0}^{\infty}\frac{(-1)^{j}}{j!}\Gamma\left(j+k+\frac{1}{2}\right)\left(\frac{\mu}{m}\right)^{2j+2k+1}\nonumber\\
&\times&\int_{0}^{1}dt\,t^{2s+2j-2}(1-t^{2})^{-s}\left[\left(\frac{m^{2}\sin^{2}\theta_0}{\mu^{2}}+t^{2}\right)^{k+\frac{1}{2}}-t^{2k+1}\right]\;.
\eea
The integral in the above expression can be evaluated in terms of hypergeometric functions, and its explicit form is
\bea\label{32}
\lefteqn{\int_{0}^{1}dt\,t^{2s+2j-2}(1-t^{2})^{-s}\left[\left(\frac{m^{2}\sin^{2}\theta_0}{\mu^{2}}+t^{2}\right)^{k+\frac{1}{2}}-t^{2k+1}\right]=\frac{\pi}{2\Gamma(s)\sin\pi s}}\nonumber\\
&&\times\left[\left(\frac{m\sin\theta_0}{\mu}\right)^{2k+1}\frac{\Gamma\left(s+j-\frac{1}{2}\right)}{\Gamma\left(j+\frac{1}{2}\right)}
{_2}F_{1}\left(-k-\frac{1}{2},s+j-\frac{1}{2},j+\frac{1}{2};-\frac{\mu^{2}}{m^{2}\sin^{2}\theta_0}\right)-\frac{\Gamma(s+j+k)}{\Gamma(j+k+1)}\right]\;,\nonumber
\eea
which is well defined in the strip $1/2-j<\Re(s)<1$. In order to be able to perform the sum over the eigenvalues $\mu$ we utilize
the Mellin-Barnes representation of the hypergeometric function \cite{gradshtein07}
\bea\label{33}
{_2}F_{1}\left(-k-\frac{1}{2},s+j-\frac{1}{2},j+\frac{1}{2};-\frac{\mu^{2}}{m^{2}\sin^{2}\theta_0}\right)=\frac{\Gamma\left(j+\frac{1}{2}\right)}{\Gamma\left(-k-\frac{1}{2}\right)\Gamma\left(s+j-\frac{1}{2}\right)}\nonumber\\
\times\frac{1}{2\pi i}\int_{\mathscr{C}} du \frac{\Gamma\left(u-k-\frac{1}{2}\right)\Gamma\left(u+s+j-\frac{1}{2}\right)\Gamma\left(-u\right)}{\Gamma\left(u+j+\frac{1}{2}\right)}\left(\frac{\mu}{m\sin\theta_0}\right)^{2u}\;,
\eea
where the contour is chosen in such a way that the poles of $\Gamma\left(u-k-\frac{1}{2}\right)\Gamma\left(u+s+j-\frac{1}{2}\right)$ lie to the left and
poles of $\Gamma(-t)$ lie to the right of $\mathscr{C}$.
%In order to be able to exchange the sum over the eigenvalues $\mu$ with the integral in (\ref{33})
%one needs to satisfy the requirement that $\Re\mathscr{C}<-1$ \cite{bordag96,kirsten01}.

\begin{figure}
\label{fig1}
   \begin{center}
    \includegraphics[scale=0.48]{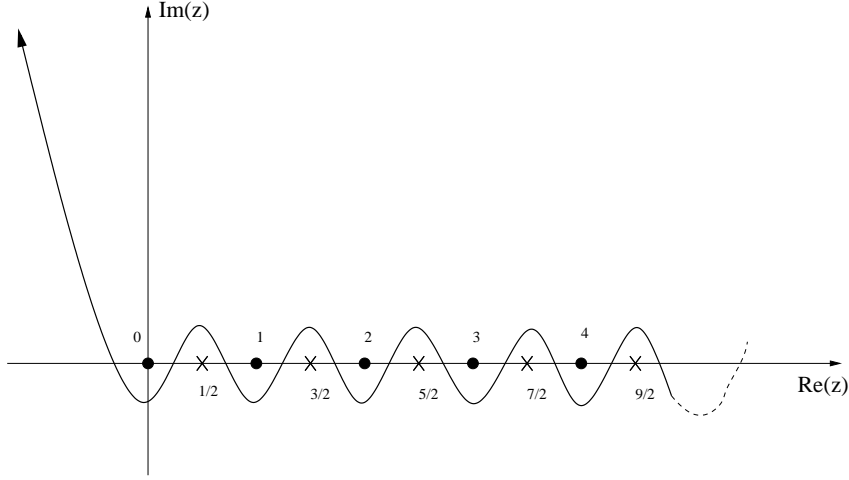}
%\vspace{.0cm}
        \caption{The figure illustrates how to deform the contour as done in passing from formula (\ref{33}) to formula (\ref{34}).}
   \end{center}
% \end{minipage}
\end{figure}

In order to proceed, we shift the contour of integration $\mathscr{C}$
so that $\Re\mathscr{C}<0$. During this procedure, the contour must cross the poles of $\Gamma\left(u-k-\frac{1}{2}\right)$ on the real axis for
$u>0$, which are positioned at $u=k+1/2$ with $k\in\mathbb{N}_{0}^{+}$ (as illustrated in Figure 1).
By taking into account the above mentioned poles and the poles to the left of the contour, one obtains
using the residue theorem
\bea
& &{_2}F_{1}\left(-k-\frac{1}{2},s+j-\frac{1}{2},j+\frac{1}{2};-\frac{\mu^{2}}{m^{2}\sin^{2}\theta_0}\right)=\frac{\Gamma\left(j+\frac{1}{2}\right)
\Gamma\left(s+k+j\right)}{\Gamma\left(s+j-\frac{1}{2}\right)\Gamma\left(k+j+1\right)}\left(\frac{\mu}{m\sin\theta_0}\right)^{2k+1}\label{34}\\
& &+\frac{\Gamma\left(j+\frac{1}{2}\right)}{\Gamma\left(-k-\frac{1}{2}\right)\Gamma\left(s+j-\frac{1}{2}\right)}\sum_{n=0}^{\infty}\frac{(-1)^{n}}{n!}
\frac{\Gamma\left(s+j+n-\frac{1}{2}\right)\Gamma\left(-s-j-k-n\right)}{\Gamma\left(-s-n+1\right)}\left(\frac{m\sin\theta_0}{\mu}\right)^{2s+2j+2n-1}.\nonumber
\eea
By substituting the last expression in formula (\ref{32}), one can prove that potentially divergent terms, in $m$ and $\sin\theta_0$, cancel exactly giving the final
result
\bea\label{35}
A_{-1}(s)&=&\frac{(\sin\theta_0)^{2s}}{4\pi\Gamma(s)}\sum_{k=0}^{\infty}\frac{(-1)^{k}(2k)!}{2^{2k}(k!)^{2}}\sum_{j=0}^{\infty}\frac{(-1)^{j}}{j!}\Gamma\left(j+k+\frac{1}{2}\right)(\sin\theta_0)^{2k+2j}\nonumber\\
&\times&\sum_{n=0}^{\infty}\frac{(-1)^{n}}{n!}\Gamma\left(s+j+n-\frac{1}{2}\right)(m\sin\theta_0)^{2n}\prod_{i=0}^{j+k}\frac{(-1)^{j+k}}{s+n+i}\,\zeta_{\mathscr{N}}\left(s+n-\frac{1}{2}\right)\;,
\eea
where in order to obtain the last formula we have used
\bea\label{35a}
\frac{\Gamma(-x-n)}{\Gamma(-x+1)}=(-1)^{n+1}\prod_{k=0}^{n}\frac{1}{x+k}\;,
\eea
which is valid for $x\neq -n$ and $n\in\mathbb{N}_{0}^{+}$, and can be proved by induction.

The next term that we need to consider is $A_{0}(s)$. The evaluation of the integral in (\ref{27}) is rather standard (see e.g. \cite{bordag96,bordag96a})
and its explicit form is
\bea\label{36}
A_{0}(s)=-\frac{(\sin\theta_{0})^{2s}}{4\Gamma(s)}\sum_{k=0}^{\infty}\frac{(-1)^{k}}{k!}\Gamma(s+k)(m\sin\theta_{0})^{2k}\zeta_{\mathscr{N}}(s+k)\;.
\eea

The last term that we want to study is $A_{i}(s)$ constructed from the functions $\Omega_{i}(\nu)$. For the computation of this integral
one needs to find a general expression for $\Omega_{i}(\nu)$ in terms of $w$. By introducing the function, defined for $i\in\mathbb{N}^{+}$,
\bea\label{36a}
\chi(i)=\frac{1+(-1)^{i}}{2}-\left[\frac{i}{2}\right]\;,
\eea
where $[x]$ represents the integer part of $x$, we can write the functions $\Omega_{i}(\nu)$ as follows
\bea\label{37}
\Omega_{i}(\nu)&=&D_{i}\left(\frac{\cos\theta_{0}}{\sqrt{1+w^{2}}}\right)\\
& &+\sum_{j=1}^{i}\frac{\sin^{2j}\theta_{0}}{(\sin^{2}\theta_{0}+w^{2})^{j}}\Bigg\{z_{0}^{(i,j)}+\sum_{b=\chi(i)}^{i}z_{i,b,j}(\cos\theta_{0})^{i+2b}(1+w^{2})^{-\frac{i}{2}-b}\Bigg\}\nonumber
\eea
The numerical coefficients $z_{0}^{(i,j)}$ and $z_{i,b,j}$ can be
obtained from the relations (\ref{21}) and (\ref{23}) and the polynomials $D_{i}(x)$ are the ones associated with the uniform asymptotic expansion of the modified Bessel
functions (see e.g \cite{bordag96,bordag96a,olver54}) defined through the cumulant expansion
\bea
\ln\left[1+\sum_{k=1}^{\infty}\frac{u_{k}(x)}{\nu^{k}}\right]\sim\sum_{n=1}^{\infty}\frac{D_{n}(x)}{\nu^{n}}\;,
\eea
where
\bea
u_{k+1}(x)=\frac{1}{2}x^{2}(1-x^{2})u_{k}^{\prime}(x)+\frac{1}{8}\int_{0}^{x}d\tau(1-5\tau^{2})u_{k}(\tau)\;,
\eea
with $u_{0}(x)=1$. They have the structure
\bea
D_{i}\left(\frac{\cos\theta_{0}}{\sqrt{1+w^{2}}}\right)=\sum_{b=0}^{i}x_{i,b}(\cos\theta_0)^{i+2b}(1+w^{2})^{-\frac{i}{2}-b}\;.
\eea
We would like to remind the reader that the functions $\Omega_{i}(\nu)$, up to the fifth order, are
listed in the appendix \ref{app1}.

By taking into account the specific form of $\Omega_{i}(\nu)$ we can write $A_{i}(s)$, for convenience, as a sum of three terms
\bea\label{38}
A_{i}(s)=I_{i,1}(s)+I_{i,2}(s)+I_{i,3}(s)\;,
\eea
with the definitions
\bea\label{39}
I_{i,1}(s)=\sum_{\mu}d(\mu)\frac{\sin\pi s}{\pi}\int_{\frac{m}{\mu}\sin\theta_{0}}^{\infty}dw\left[\frac{\mu^{2}w^{2}}{\sin^{2}\theta_{0}}-m^{2}\right]^{-s}\frac{\partial}{\partial w}\left[\frac{1}{\mu^{i}}D_{i}\left(\frac{\cos\theta_{0}}{\sqrt{1+w^{2}}}\right)\right]\;,
\eea
\bea\label{40}
I_{i,2}(s)=\sum_{\mu}d(\mu)\frac{\sin\pi s}{\pi}\sum_{j=1}^{i}\sin^{2j}\theta_{0}\int_{\frac{m}{\mu}\sin\theta_{0}}^{\infty}dw\left[\frac{\mu^{2}w^{2}}{\sin^{2}\theta_{0}}-m^{2}\right]^{-s}\frac{\partial}{\partial w}\left[\frac{z_{0}^{(i,j)}}{\mu^{i}(\sin^{2}\theta_{0}+w^{2})^{j}}\right]
\eea
and
\bea\label{41}
I_{i,3}(s)&=&\sum_{\mu}d(\mu)\frac{\sin\pi s}{\pi}\sum_{j=1}^{i}\sin^{2j}\theta_{0}\int_{\frac{m}{\mu}\sin\theta_{0}}^{\infty}dw\left[\frac{\mu^{2}w^{2}}{\sin^{2}\theta_{0}}-m^{2}\right]^{-s}\sum_{b=\chi(i)}^{i}z_{i,b,j}(\cos\theta_{0})^{i+2b}\nonumber\\
&\times&\frac{\partial}{\partial w}\left[\frac{(1+w^{2})^{-\frac{i}{2}-b}}{\mu^{i}(\sin^{2}\theta_{0}+w^{2})^{j}}\right]\;.
\eea
The first two integrals can be evaluated by following the techniques utilized in \cite{bordag96,bordag96a}. After a rather standard computation
one obtains the results
\bea\label{42}
I_{i,1}(s)&=&-\frac{(\sin\theta_{0})^{2s}}{\Gamma(s)}\sum_{j=0}^{\infty}
\frac{(-1)^{j}}{j!}(m\sin\theta_{0})^{2j}\zeta_{\mathscr{N}}\left(s+j+\frac{i}{2}\right)\\
& &\quad \quad \times \sum_{b=0}^{i}x_{i,b}(\cos\theta_0)^{i+2b}\frac{\Gamma\left(s+b+j+\frac{i}{2}\right)}{\Gamma\left(\frac{i}{2}+b\right)}\nonumber
\eea
and
\bea\label{43}
I_{i,2}(s)=-\frac{1}{\Gamma(s)}\sum_{k=0}^{\infty}\frac{(-1)^{k}}{k!}m^{2k}\zeta_{\mathscr{N}}\left(s+k+\frac{i}{2}\right)\sum_{j=1}^{i}\frac{\Gamma(s+j+k)}{\Gamma(j)}z_{0}^{(i,j)}\;.
\eea

The computation of the last integral, namely $I_{i,3}(s)$, is actually more involved. Integrating by parts and performing the change of variable $w\to m\sin\theta_0/(\mu\sqrt{t})$
one can recast (\ref{41}) in the form
\bea\label{44}
I_{i,3}(s)&=&\sum_{\mu}d(\mu)\frac{\sin\pi s}{\pi}\sum_{j=1}^{i}\sum_{b=\chi(i)}^{i}\frac{z_{i,b,j}(\cos\theta_{0})^{i+2b}}{\mu^{i}}s\,m^{-2s}\nonumber\\
&\times&\int_{0}^{1}dt\,t^{s+b+j+\frac{i}{2}-1}(1-t)^{-s-1}\left(t+\frac{m^{2}\sin^{2}\theta_0}{\mu^{2}}\right)^{-b-\frac{i}{2}}\left(t+\frac{m^{2}}{\mu^{2}}\right)^{-j}\;,
\eea
which is valid in the region $-j-b-i/2<\Re(s)<0$. By exploiting the following general integral, valid for $\Re(a)>0$ and $\Re(b)>0$, with non-vanishing real $A$ and $B$ \cite{erdelyi53,gradshtein07}
\bea\label{45}
\lefteqn{\int_{0}^{1}dx\,x^{a-1}(1-x)^{b-1}\left(x+A^{2}\right)^{-c}\left(x+B^{2}\right)^{-d}=}\nonumber\\
&&A^{-2c}B^{-2d}\frac{\Gamma(b)\Gamma(a)}{\Gamma(a+b)}F_{1}\left(a,c,d,a+b;-\frac{1}{A^{2}},-\frac{1}{B^{2}}\right)\;,
\eea
where $F_{1}$ represents the hypergeometric function of two variables, we obtain the expression
\bea\label{46}
I_{i,3}(s)&=&-\sum_{\mu}d(\mu)\sum_{j=1}^{i}\sum_{b=\chi(i)}^{i}\frac{z_{i,b,j}(\cos\theta_{0})^{i+2b}}{\mu^{i}}m^{-2s}\frac{\Gamma\left(s+b+j+\frac{i}{2}\right)}{\Gamma\left(s\right)\Gamma\left(b+j+\frac{i}{2}\right)}
\left(\frac{\mu}{m\sin\theta_0}\right)^{2b+i}\left(\frac{\mu}{m}\right)^{2j}\nonumber\\
&\times&F_{1}\left(s+b+j+\frac{i}{2},b+\frac{i}{2},j,b+j+\frac{i}{2};-\frac{\mu^{2}}{m^{2}\sin^{2}\theta_0},-\frac{\mu^{2}}{m^{2}}\right)\;.
\eea
The hypergeometric function of two variables that appears in the previous equation can be expressed in terms of a hypergeometric function
of one variable thanks to the relation \cite{erdelyi53,gradshtein07}
\bea\label{47}
F_{1}(a,b,b',b+b';x,y)=(1-x)^{-b}(1-y)^{-a+b}{_2}F_{1}\left(b+b'-a,b,b+b';\frac{y-x}{1-x}\right)\;.
\eea
The useful formula (\ref{47}) allows us to write (\ref{46}) in the form
\bea\label{47a}
I_{i,3}(s)&=&-\sum_{\mu}d(\mu)\mu^{-2s-i}\sum_{j=1}^{i}\sum_{b=\chi(i)}^{i}z_{i,b,j}(\cos\theta_{0})^{i+2b}\frac{\Gamma\left(s+b+j+\frac{i}{2}\right)}{\Gamma\left(s\right)\Gamma\left(b+j+\frac{i}{2}\right)}
\left(1+\frac{m^{2}\sin^{2}\theta_0}{\mu^{2}}\right)^{-b-\frac{i}{2}}\nonumber\\ & \times & \left(1+\frac{m^{2}}{\mu^{2}}\right)^{-j-s}
{_2}F_{1}\left(-s,b+\frac{i}{2},b+j+\frac{i}{2};\cos^{2}\theta_{0}\left(1+\frac{m^{2}\sin^{2}\theta_0}{\mu^{2}}\right)^{-1}\right)\;.
\eea
The argument of the hypergeometric function in (\ref{47a}) is smaller than one which guarantees that an expansion is applicable.
As we have done before, we utilize a Mellin-Barnes contour integral representation for the hypergeometric functions appearing in (\ref{47a}).
In this particular case, since the argument of the hypergeometric function is smaller than one, we close the Hankel contour to the right. The resulting integral can be computed with the residue theorem where the poles of the integrand are located at the positive integer points. This yields an expansion in terms of
positive powers of the parameter $m\sin\theta_0$.
On doing so, after some manipulations of the gamma functions and an expansion in powers of $m$ of the term $\left(1+m^{2}\sin^{2}\theta_0/\mu^{2}\right)^{-b-\frac{i}{2}}\left(1+m^{2}/\mu^{2}\right)^{-j-s}$, we obtain the final result for $I_{i,3}(s)$ as follows
\bea
\lefteqn{I_{i,3}(s)=-\frac{1}{\Gamma(s)}\sum_{j=1}^{i}\sum_{b=\chi(i)}^{i}z_{i,b,j}(\cos\theta_{0})^{i+2b}\frac{\Gamma\left(s+b+j+\frac{i}{2}\right)}{\Gamma\left(j+s\right)\Gamma\left(b+\frac{i}{2}\right)}
\sum_{l=0}^{\infty}\frac{(-1)^{l}}{l!}m^{2l}\Gamma(s+j+l)}\label{49}\\
&&\sum_{n=0}^{\infty}(-1)^{n}\frac{(\cos\theta_0)^{2n}}{\Gamma\left(n+b+j+\frac{i}{2}\right)}
\prod_{p=0}^{n-1}(s-p)\sum_{k=0}^{\infty}\frac{(-1)^{k}}{k!}\Gamma\left(n+k+b+\frac{i}{2}\right)(m\sin\theta_0)^{2k}\zeta_{\mathscr{N}}\left(s+l+k+\frac{i}{2}\right) ,\nonumber
\eea
where we have exploited the relation
\bea\label{w}
\frac{\Gamma(-s+n)}{\Gamma(-s)}=(-1)^{n}\prod_{p=0}^{n-1}(s-p)\;,
\eea
which is valid for $n\in\mathbb{N}^{+}$.

We would like to make a remark at this point. It is not difficult to prove that the expression for $A_{-1}(s)$, $A_{0}(s)$ and $A_{i}(s)$ derived in this section reduce, in the limit of small $\theta_{0}$, to the results obtained for the generalized cone \cite{bordag96,fucci10}.
In the case of the functions $A_{i}(s)$ the correct limiting behavior for small $\theta_0$ is achieved thanks to the following important property of $\Omega_{i}(\nu)$, namely
\begin{equation}\label{50}
z_{0}^{(i,j)}+\sum_{b=\chi(i)}^{i}z_{i,b,j}=0\;,
\end{equation}
which can be easily proved by noticing that
\begin{equation}\label{51}
\Omega_{i}(1)=D_{i}(1)\;.
\end{equation}
This last relation comes from the fact that, for $\theta_{0}\to 0$, the functions $\Omega_{i}(\nu)$ have to coincide with
the polynomials $D_{i}(\nu)$ appearing in the uniform asymptotic expansion of the modified Bessel functions.

Let us conclude this section with a few remarks about the additional complications one would encounter if $-\Delta_{\mathscr{M}}$ instead of $-\Delta_{\mathscr{M}}+d^2/4$ was considered. Formally one simply has to replace $m^2$ by $m^2 - d^2/4$. As a result, in the limit $m\to 0$ the integrals involved do not turn into doable integrals from zero to infinity, but something analytically considerably more complicated evolves. Going back to (\ref{14}) it is clear that
for $m=0$ a branch cut at $k=d/2$ would be present and when deforming the contour additional contributions due to that branch cut would arise. As stressed, because our focus here is the heat kernel where these compliations can easily be avoided, we do not discuss these additional technical complications further.

\section{The heat kernel coefficients}\label{sec5}

In this section we will use the previously obtained results for $A_{-1}(s)$, $A_{0}(s)$ and $A_{i}(s)$ in order to
evaluate the coefficients of the heat kernel asymptotic expansion on the manifold $\mathscr{M}$. Since
the base manifold $\mathscr{N}$ has been left unspecified, we will find an expression for the heat kernel coefficients
on $\mathscr{M}$ in terms of the ones on $\mathscr{N}$ \cite{bordag96}.
We will set $m=0$ in $A_{-1}(s)$, $A_{0}(s)$ and $A_{i}(s)$, because the mass dependence of the heat kernel is a trivial factor of $e^{-m^2 t}$.

For the explicit evaluation of the heat kernel coefficients associated with the operator $-\Delta_{\mathscr{M}}+d^2/4$ in arbitrary dimensions $D$, it is convenient to use the following formula
\bea\label{52}
\mathscr{A}_{\frac{n}{2}}=\Gamma\left(\frac{D-n}{2}\right)\textrm{Res}\,\zeta\left(\frac{D-n}{2}\right)\;,
\eea
which is valid for $n<D$. By keeping the dimension arbitrary, we are able to effectively compute all heat kernel coefficients.
In fact, the final formulas will be functions of the dimension $D$ of the manifold $\mathscr{M}$. Our task therefore is to evaluate $A_{-1}(s)$, $A_{0}(s)$ and $A_{i}(s)$
about $s=(D-n)/2$ in the case $m=0$. From the expressions (\ref{35}), (\ref{36}), (\ref{42}), (\ref{43}) and (\ref{49}) in the previous section, we obtain
\begin{eqnarray}\label{53}
\Gamma\left(\frac{D-n}{2}\right)\textrm{Res}\,A_{-1}\left(\frac{D-n}{2}\right)&=&\frac{(\sin\theta_0)^{D-n}}{2\sqrt{\pi}(D-n)}
\mathcal{C}_{1}(\theta_0,D-n)\nonumber\\
& &\quad \quad\times \Gamma\left(\frac{D-1-n}{2}\right)\textrm{Res}\,\zeta_{\mathscr{N}}\left(\frac{D-1-n}{2}\right)\;,
\end{eqnarray}
for the term $A_{0}(s)$ we find
\bea\label{54}
\Gamma\left(\frac{D-n}{2}\right)\textrm{Res}\,A_{0}\left(\frac{D-n}{2}\right)=-\frac{(\sin\theta_0)^{D-n}}{4}\,\Gamma\left(\frac{D-n}{2}\right)\textrm{Res}\,\zeta_{\mathscr{N}}\left(\frac{D-n}{2}\right)\;,
\eea
and, finally, for $A_{i}(s)$
\bea\label{55}
\Gamma\left(\frac{D-n}{2}\right)\textrm{Res}\,A_{i}\left(\frac{D-n}{2}\right)&=&-(\sin\theta_0)^{D-n}\Big[\mathcal{C}_{2,\,i}(\theta_0,D-n)+\mathcal{C}_{3,\,i}(\theta_0,D-n)+\mathcal{C}_{4,\,i}(\theta_0,D-n)\Big]\nonumber\\
&\times&\Gamma\left(\frac{D-n+i}{2}\right)\textrm{Res}\,\zeta_{\mathscr{N}}\left(\frac{D-n+i}{2}\right)\;.
\eea
Here we have defined the functions $\mathcal{C}_{i}(\theta_0,D-n)$ as
\bea\label{56}
\mathcal{C}_{1}(\theta_0,D-n)&=&\frac{1}{\sqrt{\pi}}\sum_{k=0}^{\infty}\frac{(2k)!}{2^{2k}(k!)^{2}}\sum_{j=0}^{\infty}
\frac{\Gamma\left(j+k+\frac{1}{2}\right)\Gamma\left(\frac{D-n-1}{2}+j\right)}{\Gamma\left(j+1\right)\Gamma\left(\frac{D-n-1}{2}\right)}\nonumber\\
& & \quad \quad \times (\sin\theta_0)^{2j+2k}\prod_{i=1}^{j+k}\frac{2}{D-n+2i}\;,
\eea
\bea\label{57}
\mathcal{C}_{2,\,i}(\theta_0,D-n)=\sum_{b=0}^{i}\tilde{x}_{i,b}\frac{\Gamma\left(\frac{D-n+i}{2}+b\right)}{\Gamma\left(\frac{D-n+i}{2}\right)\Gamma\left(b+\frac{i}{2}\right)}\;,
\eea
\bea\label{58}
\mathcal{C}_{3,\,i}(\theta_0,D-n)=(\sin\theta_0)^{n-D}\sum_{j=1}^{i}z_{0}^{(i,j)}\frac{\Gamma\left(\frac{D-n}{2}+j\right)}{\Gamma\left(\frac{D-n+i}{2}\right)\Gamma\left(j\right)}\;,
\eea
\bea\label{59}
\mathcal{C}_{4,\,i}(\theta_0,D-n)&=&(\sin\theta_0)^{n-D}\sum_{j=1}^{i}\sum_{b=\chi(i)}^{i}\tilde{z}_{i,b,j}\frac{\Gamma\left(\frac{D-n+i}{2}+b+j\right)}{\Gamma\left(\frac{D-n+i}{2}\right)\Gamma\left(b+\frac{i}{2}\right)}\nonumber\\
&\times&\sum_{n=0}^{\infty}(-1)^{n}\frac{\Gamma\left(n+b+\frac{i}{2}\right)}{\Gamma\left(n+b+j+\frac{i}{2}\right)}(\cos\theta_0)^{2n}\prod_{l=0}^{n-1}\left(\frac{D-n}{2}-l\right)\;,
\eea
with the notation
\bea
\tilde{x}_{i,b}=x_{i,b}(\cos\theta_{0})^{i+2b}\;,\qquad \tilde{z}_{i,b,j}=z_{i,b,j}(\cos\theta_{0})^{i+2b}\;.
\eea

The formulas obtained in (\ref{56}) and (\ref{59}) can actually be simplified further. In fact, the double series
which defines $\mathcal{C}_1(\theta_0,D-n)$ can be explicitly evaluated leading to the remarkably simple result
\begin{equation}\label{willy}
\mathcal{C}_{1}(\theta_0,D-n)={_2}F_{1}\left(\frac{1}{2},\frac{D-n}{2},\frac{D-n}{2}+1;\sin^2\theta_0\right)\;,
\end{equation}
which is proved in Appendix \ref{app2}. Moreover, by exploiting the
series definition of the hypergeometric function \cite{gradshtein07}
\begin{eqnarray}
{_2}F_{1}\left(\alpha,\beta,\gamma;z^2\right)=\frac{\Gamma(\gamma)}{\Gamma(\alpha)\Gamma(\beta)}\sum_{m=0}^{\infty}\frac{\Gamma(\alpha+m)\Gamma\left(\beta+m\right)}{m!\Gamma\left(\gamma+m\right)}z^{2m}
\end{eqnarray}
and the relation (\ref{w}), we can rewrite (\ref{59}) as follows
\begin{eqnarray}
\mathcal{C}_{4,\,i}(\theta_0,D-n)&=&(\sin\theta_0)^{n-D}\sum_{j=1}^{i}\sum_{b=\chi(i)}^{i}\tilde{z}_{i,b,j}\frac{\Gamma\left(\frac{D-n+i}{2}+b+j\right)}{\Gamma\left(\frac{D-n+i}{2}\right)\Gamma\left(b+j+\frac{i}{2}\right)}\nonumber\\
&\times&{_2}F_{1}\left(-\frac{D-n}{2},b+\frac{i}{2},b+j+\frac{i}{2};\cos^{2}\theta_0\right)\;.
\end{eqnarray}
The relations (\ref{53}) through (\ref{55}) are all the ingredients that we need in order to write down an expression for the heat kernel
coefficients $\mathscr{A}_{\frac{n}{2}}$ in terms of those on the base manifold. In order for $Z(s)$ in (\ref{25}) to
give no contribution to the residue of $\zeta(s)$ at the points $s=(D-n)/2$ and, in turn, to the value of the heat kernel coefficients,
we need to set $N\geq n-1$. In this way the points $s=(D-n)/2$ lie in the region of analyticity of $Z(s)$. By recalling the relation (\ref{52})
and by noticing that for the base $\mathscr{N}$ this implies
\bea\label{60}
\mathscr{A}_{\frac{n}{2}}^{\mathscr{N}}=\Gamma\left(\frac{d-n}{2}\right)\textrm{Res}\,\zeta_{\mathscr{N}}\left(\frac{d-n}{2}\right)\;,
\eea
we obtain
\bea\label{61}
\mathscr{A}_{\frac{n}{2}}&=&\frac{(\sin\theta_0)^{D-n}}{2\sqrt{\pi}(D-n)}\mathcal{C}_{1}(\theta_0,D-n)\,\mathscr{A}_{\frac{n}{2}}^{\mathscr{N}}-\frac{(\sin\theta_0)^{D-n}}{4}\mathscr{A}_{\frac{(n-1)}{2}}^{\mathscr{N}}\nonumber\\
&-&(\sin\theta_0)^{D-n}\sum_{i=1}^{n-1}\mathscr{A}_{\frac{(n-i-1)}{2}}^{\mathscr{N}}\Big[\mathcal{C}_{2,\,i}(\theta_0,D-n)+\mathcal{C}_{3,\,i}(\theta_0,D-n)+\mathcal{C}_{4,\,i}(\theta_0,D-n)\Big]\;,
\eea
where $\mathscr{A}_{\frac{n}{2}}^{\mathscr{N}}$ are the coefficients of the small $t$ asymptotic expansion of the operator $-\Delta_{\mathscr{N}}+(d-1)^{2}/4$ on $\mathscr{N}$, and  $\mathscr{A}_{\frac{(n-1)}{2}}^{\mathscr{N}}=0$ for $n=0$. The coefficients $\mathcal{A}^{\mathscr{M}}_{\frac{n}{2}}$ of the heat kernel asymptotic expansion for the Laplacian on
the spherical suspension $\mathscr{M}$ are obtained from $\mathscr{A}_{\frac{n}{2}}$ using the relation
\begin{equation}\label{61a}
 \mathcal{A}^{\mathscr{M}}_{\frac{n}{2}}=\sum_{k=0}^{\left[\frac{n}{2}\right]}\frac{1}{k!}\left(\frac{d}{2}\right)^{2k}\mathscr{A}_{\frac{n}{2}-k}\;.
\end{equation}
The last formula, together with the explicit expression for $\mathscr{A}_{\frac{n}{2}}$ in (\ref{61}), represents the {\it main result} of our paper. It provides a relation for the heat kernel coefficients
on the spherical suspension $\mathscr{M}$ in terms of the heat kernel coefficients of the operator $-\Delta_{\mathscr{N}}+(d-1)^{2}/4$ for any smooth, possibly with boundary, base $\mathscr{N}$. This result is very general
considering the fact that it gives any desired heat kernel coefficient for arbitrary dimensions $D$. In the leading order, the expression for $\mathcal{A}_{\frac{n}{2}}^{\mathscr{M}}$ in (\ref{61}) reduces to the one for the heat kernel asymptotic coefficients on the generalized cone
once the limit of small $\theta_0$ is considered \cite{bordag96}.
One can obtain more explicit formulas once $\mathscr{N}$ is specified.

One might ask how the complicated $\theta_0$ dependence in terms of hypergeometric functions in (\ref{61}) results from the known expressions of the heat kernel coefficients in terms of geometric tensors. This is in fact a direct consequence of integrating trigonometric functions raised to arbitrary powers. For example the leading coefficient $${\mathcal{A}}^{\mathscr{M}} _0 = (4\pi)^{-D/2} \mbox{vol} ({\mathscr{M}}).$$ For the given manifold, the volume is $$\mbox{vol} ({\mathscr{M}}) = \mbox{vol} ({\mathscr{N}}) \,\, \int\limits_0^{\theta_0} d\theta \sin ^{D-1} \theta = \mbox{vol} ({\mathscr{N}}) \,\, \frac{\sin \theta_0} D \mathcal{C} _1 (\theta_0 , D).$$ Similarly, integrating quantities involving curvature terms like for example the extrinsic curvature $$K = d \cot \theta_0$$ or the Riemann scalar curvature $$R = \frac 1 {\sin ^2 \theta} R_{\mathscr{N}} + \frac{ d (1-d)} {\sin ^2\theta} + d (1+d) $$
and powers thereof,
results obtained from (\ref{61a}) can be reproduced. Evaluating the heat kernel coefficients from the curvature tensors is actually quite cumbersome and only possible for the leading known coefficients \cite{gilkey95,gilkey04,kirsten01}. For the given geometry, (\ref{61a}) has the advantage that any number of coefficients can be determined using a simple algebraic computer program. In principle this could be used to put
restrictions on higher heat kernel coefficients, in the same way that the generalized cone was used \cite{bordag96,kirs98-15-5,dowk01-42-434}.

In the next section, we will study the situation in which
the base manifold is a $d$-dimensional sphere. For this particular choice the spectral zeta function $\zeta_{\mathscr{N}}(s)$ can be written
in terms of a linear combination of Barnes zeta functions and more specific results can be obtained \cite{bordag96,chang93}.

\section{A specific Base Manifold}\label{sec6}

In this section we will focus our attention to a particular case for which more explicit results for the heat kernel coefficients can be obtained .
We consider the base manifold $\mathscr{N}$ to be a $d$-dimensional sphere of unit radius. The eigenvalues of the modified Laplacian
$-\Delta_{\mathscr{N}}+(d-1)^{2}/4$ are known to be
\bea\label{62}
\mu^{2}=\left(k+\frac{d-1}{2}\right)^{2}\;,
\eea
where $k\geq 0$, and the eigenfunctions are hyperspherical harmonics with degeneracy
\bea\label{63}
d( k)=(2k+d-1)\frac{(k+d-2)!}{k!(d-1)!}\;.
\eea
The two formulas above, together with the definition of the base zeta function $\zeta_{\mathscr{N}}(s)$ in (\ref{12}), allow us
to write
\bea\label{64}
\zeta_{{\mathscr{N}}}(s)=\sum_{k=0}^{\infty}(2k+d-1)\frac{(k+d-2)!}{k!(d-1)!}\left(k+\frac{d-1}{2}\right)^{-2s}\;.
\eea
At this point, it is convenient to introduce the Barnes zeta function, which is defined as follows \cite{barnes,dowker05}
\bea\label{65muldef}
\zeta_{{\cal B}}(s,a|\vec{r})=\sum_{\vec{m}=0}^{\infty}(a+\vec{m}\cdot\vec{r})^{-s}\;,
\eea
valid for $\Re(s)>d$, where $\vec{m}$ and $\vec{r}$ are $d$-dimensional vectors.
It has been shown in \cite{bordag96,chang93} that $\zeta_{\mathscr{N}}(s)$ in (\ref{64}) can be rewritten, after some algebraic manipulations,
as a sum of two Barnes zeta functions
\bea\label{65}
\zeta_{{\mathscr{N}}} (s) =\zeta_{\mathcal{B}}\left(2s,\frac{d+1}{2}\right)+\zeta_{\mathcal{B}}\left(2s,\frac{d-1}{2}\right)\;,
\eea
where we have used the notation $\zeta_{{\cal B}} (s,a|\vec 1)=\zeta_{{\cal B}} (s,a)$.

For the purpose of the evaluation of the heat kernel coefficients we will need the residues of the function $\zeta_{\mathscr{N}}(s)$ at $s=m/2$ with $m\in\mathbb{N}^{+}$.
To this end it is convenient to utilize the following integral representation of the Barnes zeta function \cite{bordag96,fucci10}
\bea\label{66}
\zeta_{{\cal B}}(s,a)=\frac{i\Gamma(1-s)}{2\pi}\int_{L}dy\,\frac{e^{y\left(\frac{d}{2}-a\right)}(-y)^{s-1}}{2^{d}\sinh^{d}\left(\frac{y}{2}\right)}\;,
\eea
where $L$ represents the Hankel contour. The spectral zeta function on the base manifold can now be represented in terms of a contour integral
by using the relation (\ref{65}) and the formula (\ref{66}). More explicitly, one obtains \cite{bordag96}
\bea\label{67}
\zeta_{{\mathscr{N}}}(s)=\frac{i\Gamma(2-2s)}{2\pi (d-1)} 2^{2s+1-d}\,\,\int_{L}dy\,(-y)^{2s-2} \frac 1 {\sinh^{d-1} y }\;.
\eea
%The desired expression is achieved by making a change of variables, $y/2\to y$, which recasts the above relation
%in the form \cite{bordag96,fucci10}
%\bea\label{68}
%\zeta_{{\mathscr{N}}}(s)=(-1)^{2s-2}\frac{i\Gamma(2-2s)}{2\pi(d-1)}2^{2s+1-d}\sum_{\nu=0}^{\infty}\frac{D_{\nu}^{(d-1)}}{\nu!}\int_{L}dy\,y^{2s-d-1+\nu}\;,
%\eea
%where the coefficients $D_{\nu}^{(d-1)}$ can be easily computed, by equating like powers of $y$, from the formula
%\bea\label{69}
%\left(\frac{y}{\sinh y}\right)^{d-1}=\sum_{\nu=0}^{\infty}D_{\nu}^{(d-1)}\frac{y^{\nu}}{\nu!}\;.
%\eea
%According to the relation (\ref{60}), we need the residue of $\zeta_{{\mathscr{N}}}(s)$ at the points $s=m/2$ with $m$ being a
%positive integer. It is not very difficult to see that the contour integral in (\ref{68}) vanishes unless $\nu=d-m$ in which case the integrand
%has a simple pole. By using the residue theorem we get \cite{bordag96,fucci10}
%\bea\label{70}
%\textrm{Res}\,\zeta_{{\mathscr{N}}}\left(\frac{m}{2}\right)=\frac{2^{m-d}D_{d-m}^{(d-1)}}{(d-1)(m-2)!(d-m)!}\;,
%\eea
%that is well defined for $m\geq 2$ and $d\geq m$. We would like to point out that $D_{d-m}^{(d-1)}$ will give
%non-vanishing contributions only if the quantity $(d-m)$ is even as one can easily see by recalling (\ref{69}).
The residues at $s=m/2$ for $m\geq 2$ and $d\geq m$ are then easily found by an application of the residue theorem. They are best expressed in terms
of $D_\nu^{(d-1)}$ defined through
\bea\label{69}
\left(\frac{y}{\sinh y}\right)^{d-1}=\sum_{\nu=0}^{\infty}D_{\nu}^{(d-1)}\frac{y^{\nu}}{\nu!}\;.
\eea
In detail one obtains
\bea\label{70}
\textrm{Res}\,\zeta_{{\mathscr{N}}}\left(\frac{m}{2}\right)=\frac{2^{m-d}D_{d-m}^{(d-1)}}{(d-1)(m-2)!(d-m)!}\;.
\eea
This result can now be used in (\ref{60})
in order to get a formula for the heat kernel coefficients $\mathscr{A}_{\frac{n}{2}}^{\mathscr{N}}$.
One obtains
\bea\label{71}
\mathscr{A}_{\frac{n}{2}}^{\mathscr{N}}=\frac{(d-n-1)D_{n}^{(d-1)}}{2^{n}n!(d-1)\Gamma(d-n)}\Gamma\left(\frac{d-n}{2}\right)\;.
\eea
This expression can be written in terms of a single gamma function by exploiting the formula
\bea\label{72}
\Gamma(d-n)=(2\pi)^{-\frac{1}{2}}2^{d-n-\frac{1}{2}}\Gamma\left(\frac{d-n+1}{2}\right)\Gamma\left(\frac{d-n}{2}\right)\;,
\eea
which is a particular case of the more general product theorem for gamma functions \cite{gradshtein07}. This last remark allows us to write (\ref{71}) as
\bea\label{73}
\mathscr{A}_{\frac{n}{2}}^{\mathscr{N}}=2\sqrt{\pi}\,\frac{(d-n-1)D_{n}^{(d-1)}}{2^{d}n!(d-1)\Gamma\left(\frac{d-n+1}{2}\right)}\;.
\eea
This result, applied to the general relation (\ref{60}), gives an explicit formula for the heat kernel coefficients
$\mathscr{A}_{\frac{n}{2}}$ in the situation in which the base is a $d$-dimensional sphere. In more detail we have
\bea\label{74}
\frac{(4\pi)^{\frac{D}{2}}}{|S_{d}|}\mathscr{A}_{\frac{n}{2}}&=&\frac{(\sin\theta_0)^{D-n}(d-n-1)}{n!(d-1)(d-n+1)}\left(\frac{d-n+1}{2}\right)_{\frac{n}{2}}D_{n}^{(d-1)}\mathcal{C}_{1}(\theta_0,D-n)\nonumber\\
&-&\frac{\sqrt{\pi}\,(\sin\theta_0)^{D-n}(d-n)}{2(d-1)(n-1)!}\left(\frac{d-n+2}{2}\right)_{\frac{n-1}{2}}D_{n-1}^{(d-1)}\nonumber\\
&-&\frac{2\sqrt{\pi}}{(d-1)}(\sin\theta_0)^{D-n}\sum_{i=1}^{n-1}\frac{(d-n+i)}{(n-1-i)!}\left(\frac{d-n+i+2}{2}\right)_{\frac{n-i-1}{2}}D_{n-i-1}^{(d-1)}\nonumber\\
&\times&\Big[\mathcal{C}_{2,\,i}(\theta_0,D-n)+\mathcal{C}_{3,\,i}(\theta_0,D-n)+\mathcal{C}_{4,\,i}(\theta_0,D-n)\Big]\;,
\eea
where we have used the Pochhammer symbol $(x)_{n}=\Gamma(x+n)/\Gamma(x)$ and $|S_{d}|$ represents the surface area of a
$d$-dimensional sphere of unit radius
\bea
|S_{d}|=\frac{(4\pi)^{\frac{d}{2}}}{(d-1)!}\Gamma\left(\frac{d}{2}\right)\;.
\eea
We would like to stress, one more time, that the result (\ref{74}) reduces, in the leading order, to the one obtained in \cite{bordag96} once the limit
of small angles $\theta_0$ is considered. The formula (\ref{74}) allows for a quick computation of all the heat kernel coefficients of the manifold $\mathscr{M}$, through the relation (\ref{61}), in arbitrary dimensions in the case in which the base manifold is a $d$-dimensional sphere. The resulting expressions will depend in a direct way on
the functions $\mathcal{C}_{i}$ introduced in (\ref{56}) through (\ref{59}).

Below, we will list the heat kernel coefficients up to $\mathcal{A}_{3}^{\mathscr{M}}$ for a $d$-dimensional sphere as the base manifold. By utilizing (\ref{61a}), (\ref{74}) and by introducing the function
\begin{equation}\label{75}
  \mathcal{F}_{i}(\theta_0,D-n)=\mathcal{C}_{2,\,i}(\theta_0,D-n)+\mathcal{C}_{3,\,i}(\theta_0,D-n)+\mathcal{C}_{4,\,i}(\theta_0,D-n)\;,
\end{equation}
we have
\bea
\frac{(4\pi)^{\frac{D}{2}}}{(\sin\theta_0)^{1+d} |S_{d}|}\mathcal{A}_{0}^{\mathscr{M}}=\frac{1}{d+1}\,\mathcal{C}_{1}(\theta_0,D)\;,
\eea
\bea
\frac{(4\pi)^{\frac{d}{2}}}{(\sin\theta_0)^{d}|S_{d}|}\mathcal{A}_{\frac{1}{2}}^{\mathscr{M}}=-\frac{1}{4}\;,
\eea
\bea
\frac{(4\pi)^{\frac{D}{2}}}{(\sin\theta_0)^{d-1} |S_{d}|}\mathcal{A}_{1}^{\mathscr{M}}&=&-\frac{d-3}{12}\,\mathcal{C}_{1}(\theta_0,D-2)-2 \sqrt{\pi }\,\mathcal{F}_{1}(\theta_0,D-2)\nonumber\\
&+&\frac{d^{2}}{4(d+1)}\sin^{2}\theta_{0}C_{1}(\theta_0,D)\;,
\eea
\bea
\frac{(4\pi)^{\frac{d}{2}}}{(\sin\theta_0)^{d-2} |S_{d}|}\mathcal{A}_{\frac{3}{2}}^{\mathscr{M}}&=&\frac{(d-3) (d-1)}{48}- \mathcal{F}_{2}(\theta_0,D-3)-\frac{d^{2}}{16}\sin^{2}\theta_{0}\;,
\eea
\bea
\frac{(4\pi)^{\frac{D}{2}}}{(\sin\theta_0)^{d-3} |S_{d}|}\mathcal{A}_{2}^{\mathscr{M}}&=&\frac{(d-5) (d-1) (5d-3)}{1440} \mathcal{C}_{1}(\theta_0,D-4)+\sqrt{\pi}\,\frac{(d-3)(d-1)}{6}\mathcal{F}_{1}(\theta_0,D-4)\nonumber\\
&-&2\sqrt{\pi}\,\mathcal{F}_{3}(\theta_0,D-4)-\frac{d^{2}(d-3)}{48}\sin^{2}\theta_0\,\mathcal{C}_{1}(\theta_0,D-2)\nonumber\\
&-&\frac{d^{2}}{2}\sqrt{\pi}\sin^{2}\theta_0\,\mathcal{F}_{1}(\theta_0,D-2)+\frac{d^{4}}{32(d+1)}\sin^{4}\theta_0 \mathcal{C}_{1}(\theta_0,D) \;,
\eea
\bea
\frac{(4\pi)^{\frac{d}{2}}}{(\sin\theta_0)^{d-4} |S_{d}|}\mathcal{A}_{\frac{5}{2}}^{\mathscr{M}}&=&-\frac{(d-5) (d-3) (d-1) (5d-3)}{5760}+\frac{(d-3)(d-1)}{12}\mathcal{F}_{2}(\theta_0,D-5)\nonumber\\
&-&\mathcal{F}_{4}(\theta_0,D-5)+\frac{d^{2}(d-3)(d-1)}{192}\sin^{2}\theta_0-\frac{d^{2}}{4}\sin^{2}\theta_0\mathcal{F}_{2}(\theta_0,D-3)\nonumber\\
&-&\frac{d^{4}}{128}\sin^{4}\theta_0\;,
\eea
\bea
\frac{(4\pi)^{\frac{D}{2}}}{(\sin\theta_0)^{d-5} |S_{d}|}\mathcal{A}_{3}^{\mathscr{M}}&=&-\frac{(d-7) (d-3) (d-1) \left(9-28 d+35 d^2\right)}{362880}\mathcal{C}_{1}(\theta_0,D-6)\nonumber\\
&-& \sqrt{\pi } \,\frac{(d-5) (d-3) (d-1) (5d-3)}{720}\mathcal{F}_{1}(\theta_0,D-6)\nonumber\\
&+& \sqrt{\pi }\frac{(d-3) (d-1)}{6}\mathcal{F}_{3}(\theta_0,D-6)-2 \sqrt{\pi }\,\mathcal{F}_{5}(\theta_0,D-6)\nonumber\\
&+&\frac{d^{2}(d-5)(d-1)(5d-3)}{5760}\sin^{2}\theta_0\mathcal{C}_{1}(\theta_0,D-4)\nonumber\\
&+&\frac{d^{2}(d-3)(d-1)}{24}\sqrt{\pi}\sin^{2}\theta_0\mathcal{F}_{1}(\theta_0,D-4)-\frac{d^{2}}{2}\sqrt{\pi}\sin^{2}\theta_0 \mathcal{F}_{3}(\theta_0,D-4)\nonumber\\
&-&\frac{d^{4}(d-3)}{384}\sin^{4}\theta_0 \mathcal{C}_{1}(\theta_0,D-2)-\frac{d^{4}}{16}\sqrt{\pi}\sin^{4}\theta_0 \mathcal{F}_{1}(\theta_0,D-2)\nonumber\\
&+&\frac{d^{6}}{384(d+1)}\sin^{6}\theta_0 \mathcal{C}_{1}(\theta_0,D)\;.
\eea
For reasons of space, the above results are expressed in terms of the functions $\mathcal{F}_{i}(\theta_0,D-n)$ which are defined in equation (\ref{75}) together with (\ref{57}) through (\ref{59}). We would like to stress, once again, that one can easily evaluate the functions $\mathcal{F}_{i}(\theta_0,D-n)$ explicitly with the help of a simple computer
program.

\section{Conclusions}

In this paper we have analyzed the spectral zeta function for a Laplace operator acting on scalar functions defined on the spherical suspension and evaluated
the coefficients of the asymptotic expansion of the heat kernel for Dirichlet boundary conditions. The coefficients were given
in terms of those on the base. More explicit result were presented once the base manifold was specified to be a
$d$-dimensional sphere. We have also provided a systematic way of dealing with the integrals defining $A_{-1}(s)$, $A_{0}(s)$ and $A_{i}(s)$
coming from the uniform asymptotic expansion of the Legendre functions used for the analytic continuation of the spectral zeta function.
These represent non trivial results because they cannot be obtained by direct and straightforward application of previous techniques utilized when the radial eigenfunctions
are Bessel functions.

Since in this work we have only treated the case in which the Laplace operator is endowed with Dirichlet boundary conditions, the next
natural step would be to extend the results to the more general Robin case. This analysis is possible because
the uniform asymptotic expansion of the derivative of the Legendre functions is known
\cite{khusnutdinov03,thorne57} and it has the same form as (\ref{17}). We expect that the analytic continuation of the spectral zeta function
for Robin boundary conditions would follow the lines presented in this paper without any major technical complications.

It would also be very interesting to apply the formalism developed in this paper to the evaluation of the Casimir energy
for the spherical suspension following the ideas developed for conical manifolds in \cite{fucci11,fucci11b}. It is well known that the regularized vacuum energy is obtained from the value
of the spectral zeta function at the point $s=-1/2$. One would need to analyze in detail the term $Z(s)$ in (\ref{25}). While for the computation of the heat kernel coefficients this term
does not contribute to the residues of $\zeta(s)$, we would be forced to deal with it in the evaluation of the Casimir energy.
One way of expressing $Z(s)$ would be by means of the Abel-Plana summation formula which has been used
in \cite{flachi10}. The results obtained in this way would be useful, at least, for a numerical estimate of
the Casimir energy on the spherical suspension.

It is not clear, at this moment, whether or not the results
obtained in this work can be easily adapted to the evaluation of the spectral zeta function associated with $-\Delta_{\mathscr{M}}$. Although the eigenfunctions
are still associated Legendre functions, this case leads to a more cumbersome uniform asymptotic expansion
(the polynomials $\Omega_{n}(\nu)$ would acquire non-polynomials terms containing the arcsine \cite{khusnutdinov03}) and new technicalities have to be considered.

\begin{acknowledgments}
KK is supported by the National Science Foundation Grant
PHY-0757791.
\end{acknowledgments}

\begin{appendix}

\section{The polynomials $\Omega_{n}(\nu)$}\label{app1}

In this appendix we list the functions $\Omega_{n}(\nu)$ for $n$ up to five. In the following we will set
$\gamma=w/\sin\theta_0$ and $\nu$ is defined in (\ref{18}). By utilizing the recurrence relation (\ref{21}) and the
cumulant expansion (\ref{23}) we get
\bea
\Omega_{1}(\nu)=\frac{\nu}{8}-\frac{5 \nu^3}{24}+\frac{1 }{ \left(1+\gamma ^2\right)}\left(\frac{1}{24}-\frac{\nu}{4}+\frac{5 \nu^3}{24}\right)\;,
\eea
\bea
\Omega_{2}(\nu)&=&\frac{\nu^2}{16}-\frac{3 \nu^4}{8}+\frac{5 \nu^6}{16}+\frac{1}{\left(1+\gamma ^2\right)^2}\left(-\frac{1}{8}+\frac{9 \nu^2}{16}-\frac{3 \nu^4}{4}+\frac{5 \nu^6}{16}\right)\nonumber\\
&+&\frac{1}{\left(1+\gamma ^2\right)}\left(\frac{1}{16}-\frac{9 \nu^2}{16}+\frac{9 \nu^4}{8}-\frac{5 \nu^6}{8}\right)\;,
\eea
\bea
\Omega_{3}(\nu)&=&\frac{25 \nu^3}{384}-\frac{531 \nu^5}{640}+\frac{221 \nu^7}{128}-\frac{1105 \nu^9}{1152}\nonumber\\
&+&\frac{1}{ \left(1+\gamma ^2\right)^3}\left(-\frac{7}{720}+\frac{19 \nu}{32}-\frac{259 \nu^3}{96}+\frac{2949 \nu^5}{640}-\frac{221 \nu^7}{64}+\frac{1105 \nu^9}{1152}\right)\nonumber\\
&+&\frac{1}{\left(1+\gamma ^2\right)^2}\left(\frac{7}{960}-\frac{19 \nu}{32}+\frac{259 \nu^3}{64}-\frac{2949 \nu^5}{320}+\frac{1105 \nu^7}{128}-\frac{1105 \nu^9}{384}\right)\nonumber\\
&+&\frac{1 }{ \left(1+\gamma ^2\right)}\left(\frac{9 \nu}{128}-\frac{71 \nu^3}{48}+\frac{87 \nu^5}{16}-\frac{221 \nu^7}{32}+\frac{1105 \nu^9}{384}\right)\;,
\eea
\bea
\Omega_{4}(\nu)&=&\frac{13 \nu^4}{128}-\frac{71 \nu^6}{32}+\frac{531 \nu^8}{64}-\frac{339 \nu^{10}}{32}+\frac{565 \nu^{12}}{128}\nonumber\\
&+&\frac{1}{\left(1+\gamma ^2\right)^4}\left(\frac{5}{16}-\frac{153 \nu^2}{32}+\frac{2613 \nu^4}{128}-\frac{1283 \nu^6}{32}+\frac{2619 \nu^8}{64}-\frac{339 \nu^{10}}{16}+\frac{565 \nu^{12}}{128}\right)\nonumber\\
&+&\frac{1}{\left(1+\gamma ^2\right)^3}\left(-\frac{5}{16}+\frac{459 \nu^2}{64}-\frac{2613 \nu^4}{64}+\frac{6415 \nu^6}{64}-\frac{7857 \nu^8}{64}+\frac{2373 \nu^{10}}{32}-\frac{565 \nu^{12}}{32}\right)\nonumber\\
&+&\frac{1}{\left(1+\gamma ^2\right)^2}\left(\frac{5}{128}-\frac{171 \nu^2}{64}+\frac{3207 \nu^4}{128}-\frac{677 \nu^6}{8}+\frac{2097 \nu^8}{16}-\frac{3051 \nu^{10}}{32}+\frac{1695 \nu^{12}}{64}\right)\nonumber\\
&+&\frac{1}{ \left(1+\gamma ^2\right)}\left(\frac{9 \nu^2}{64}-\frac{297 \nu^4}{64}+\frac{1709 \nu^6}{64}-\frac{3681 \nu^8}{64}+\frac{1695 \nu^{10}}{32}-\frac{565 \nu^{12}}{32}\right)\;,
\eea
\bea
\Omega_{5}(\nu)&=&\frac{1073 \nu^5}{5120}-\frac{50049 \nu^7}{7168}+\frac{186821 \nu^9}{4608}-\frac{44899 \nu^{11}}{512}+\frac{82825 \nu^{13}}{1024}-\frac{82825 \nu^{15}}{3072}\nonumber\\
&+&\frac{1}{\left(1+\gamma ^2\right)^5}\Bigg(\frac{31}{2520}-\frac{631 \nu}{128}+\frac{39551 \nu^3}{768}-\frac{539643 \nu^5}{2560}+\frac{3250189 \nu^7}{7168}-\frac{162353 \nu^9}{288}\nonumber\\
&+&\frac{209571 \nu^{11}}{512}-\frac{82825 \nu^{13}}{512}+\frac{82825 \nu^{15}}{3072}\Bigg)\nonumber\\
&+&\frac{1}{ \left(1+\gamma ^2\right)^4}\Bigg(-\frac{31}{2016}+\frac{1893 \nu}{256}-\frac{39551 \nu^3}{384}+\frac{539643 \nu^5}{1024}-\frac{9750567 \nu^7}{7168}+\frac{1136471 \nu^9}{576}\nonumber\\
&-&\frac{209571 \nu^{11}}{128}+\frac{745425 \nu^{13}}{1024}-\frac{414125 \nu^{15}}{3072}\Bigg)\nonumber
\eea
\bea
&+&\frac{1}{\left(1+\gamma ^2\right)^3}\Bigg(\frac{31}{8064}-\frac{1415 \nu}{512}+\frac{65569 \nu^3}{1024}-\frac{116327 \nu^5}{256}+\frac{10693979 \nu^7}{7168}-\frac{6031529 \nu^9}{2304}\nonumber\\
&+&\frac{1302325 \nu^{11}}{512}-\frac{82825 \nu^{13}}{64}+\frac{414125 \nu^{15}}{1536}\Bigg)\nonumber\\
&+&\frac{1}{\left(1+\gamma ^2\right)^2}\Bigg(\frac{153 \nu}{1024}-\frac{38503 \nu^3}{3072}+\frac{158319 \nu^5}{1024}-\frac{733859 \nu^7}{1024}+\frac{2476075 \nu^9}{1536}-\frac{972981 \nu^{11}}{512}\nonumber\\
&+&\frac{579775 \nu^{13}}{512}-\frac{414125 \nu^{15}}{1536}\Bigg)\nonumber\\
&+&\frac{1}{ \left(1+\gamma ^2\right)}\Bigg(\frac{183 \nu^3}{512}-\frac{8613 \nu^5}{512}+\frac{141923 \nu^7}{1024}-\frac{170509 \nu^9}{384}+\frac{86067 \nu^{11}}{128}\nonumber\\
&-&\frac{248475 \nu^{13}}{512}+\frac{414125 \nu^{15}}{3072}\Bigg)\;.
\eea

\section{The function $\mathcal{C}_{1}(\theta_0,D-n)$}\label{app2}

In this appendix we will provide a proof of the formula (\ref{willy}). Let us start with the expression for $\mathcal{C}_{1}$
in terms of the double infinite series (\ref{56}),
\bea\label{appendix1}
\mathcal{C}_{1}(\theta_0,2s)=\frac{1}{\sqrt{\pi}}\sum_{k=0}^{\infty}\frac{(2k)!}{2^{2k}(k!)^{2}}\sum_{j=0}^{\infty}
\frac{\Gamma\left(j+k+\frac{1}{2}\right)\Gamma\left(s+j-\frac{1}{2}\right)\Gamma(s+1)}{\Gamma\left(j+1\right)\Gamma\left(s-\frac{1}{2}\right)\Gamma(s+1+j+k)}(\sin\theta_0)^{2j+2k}\;,
\eea
where we have used the fact that \cite{gradshtein07}
\bea
\prod_{i=1}^{j+k}\frac{1}{s+i}=\frac{\Gamma(s+1)}{\Gamma(s+1+j+k)}\;.
\eea
In addition, one can prove by a direct calculation and by using the doubling formula for the Gamma function \cite{gradshtein07}
\bea
\Gamma(2k)=2^{2k-1}\frac{\Gamma(k)\Gamma\left(k+\frac{1}{2}\right)}{\Gamma\left(\frac{1}{2}\right)}\;,
\eea
that
\bea
\frac{(2k)!}{2^{2k}k!}=\frac{2k\Gamma(2k)}{2^{2k}k\Gamma(k)}=\frac{\Gamma\left(k+\frac{1}{2}\right)}{\Gamma\left(\frac{1}{2}\right)}\;.
\eea

The previous results allow us to rewrite the double sum in (\ref{appendix1}) as
\bea\label{appendix2}
\mathcal{C}_{1}(\theta_0,2s)=\frac{1}{\sqrt{\pi}}\sum_{k=0}^{\infty}\sum_{j=0}^{\infty}\frac{\Gamma\left(j+k+\frac{1}{2}\right)\Gamma\left(s+j-\frac{1}{2}\right)\Gamma\left(k+\frac{1}{2}\right)\Gamma(s+1)}
{k!j!\,\Gamma\left(s-\frac{1}{2}\right)\Gamma\left(\frac{1}{2}\right)\Gamma(s+1+j+k)}(\sin\theta_0)^{2j+2k}\;.
\eea
By directly comparing the last expression with the definition of the hypergeometric function of two variables \cite{gradshtein07}
\bea\label{appendix3}
F_{1}(\alpha,\beta,\beta',\gamma;x,y)=\frac{\Gamma(\gamma)}{\Gamma(\alpha)\Gamma(\beta)\Gamma(\beta')}\sum_{m=0}^{\infty}\sum_{n=0}^{\infty}\frac{\Gamma(\alpha+m+n)\Gamma(\beta+m)\Gamma(\beta'+n)}{m!n!\Gamma(\gamma+m+n)}x^{m}y^{n}\;,
\eea
and by recalling that $\Gamma(1/2)=\sqrt{\pi}$, one finds that
\bea
\mathcal{C}_{1}(\theta_0,2s)=F_{1}\left(\frac{1}{2},s-\frac{1}{2},\frac{1}{2},s+1;\sin^2\theta_0,\sin^2\theta_0\right)\;.
\eea
The above hypergeometric function of two variables can be written in terms of a hypergeometric function
of one variable thanks to the relation \cite{gradshtein07}
\bea
F_{1}(\alpha,\beta,\beta',\gamma;x,x)={_2}F_{1}(\alpha,\beta+\beta',\gamma;x)\;,
\eea
which leads to the formula (\ref{willy}), once $s=(D-n)/2$ has been put.

\end{appendix}

\end{document}